\documentclass[twocolumn,aps]{revtex4-1}
\usepackage{graphicx}
\usepackage{float}

\begin{document}

\title{Improvement of the 3$\omega$ thermal conductivity measurement technique at nanoscale}

\author{G.Pennelli}
\email{g.pennelli@iet.unipi.it, tel. +39 050 2217 699, fax. +39 050 2217522}
\affiliation{Dipartimento di Ingegneria della Informazione, Universit\`a di Pisa, Via G.Caruso, I-56122 Pisa, Italy}

\author{E.Dimaggio}
\affiliation{Dipartimento di Ingegneria della Informazione, Universit\`a di Pisa, Via G.Caruso, I-56122 Pisa, Italy}

\author{M.Macucci}
\affiliation{Dipartimento di Ingegneria della Informazione, Universit\`a di Pisa, Via G.Caruso, I-56122 Pisa, Italy}

\begin{abstract}
The reduction of the thermal conductivity in nanostructures opens up the 
possibility of exploiting for thermoelectric purposes also materials such 
as silicon, which are cheap, available and sustainable but with a 
high thermal conductivity in their bulk form. 
The development of thermoelectric devices based on these innovative 
materials requires reliable techniques for the measurement 
of thermal conductivity on a nanometric scale. 
The approximations introduced by conventional techniques for 
thermal conductivity measurements can lead to unreliable results when applied 
to nanostructures, because heaters and temperature sensors needed for the 
measurement cannot have a negligible size, and therefore perturb the result. 
In this paper we focus on the 3$\omega$ technique, applied to the thermal 
conductivity measurement of suspended silicon nanomembranes. 
To overcome the approximations introduced by conventional 
analytical models used for the interpretation of the 3$\omega$ data, we propose
to use a numerical solution, performed by means of finite element modeling, 
of the thermal and 
electrical transport equations. An excellent fit of the experimental data 
will be presented, discussed, and compared with an analytical model.
\end{abstract}

\maketitle

\section{Introduction}
Thermoelectric applications require the development of materials with a large
value of the figure of merit $ZT=S^2 \sigma /k_t~T$, where $S$ is the Seebeck 
coefficient, $\sigma$ is the electrical 
conductivity, $k_t$ is the thermal conductivity and $T$ is the absolute 
temperature.
Recently, it has been demonstrated that $k_t$ is strongly reduced in 
nanostructures, such
as nanowires\cite{heat-2003,heat-2008,park-2011, mio-conduzionetermica}, where 
the phonon propagation is limited by scattering on the nanowire walls. 
Interesting results in rough nanowires\cite{lim-2012, feser-2012}, where the 
effect of phonon scattering on the surfaces
is increased, open interesting perspectives for the fabrication of efficient 
thermoelectric 
generators to be used for energy recovery and/or green-energy harvesting. 
Nanostructuring should allow the fabrication of thermoelectric generators 
based on materials, such as silicon, 
which are cheap, sustainable, very stable over a large range of temperatures, 
but which have a high thermal conductivity in their bulk state
($k_t=150$ W/mK for bulk silicon).
The development of nanostructured materials and thermoelectric devices 
requires the 
improvement of existing techniques for the measurement of the thermal 
conductivity, because the size of both the heaters and the temperature 
sensors needed 
for determining $k_t$ cannot in practice be much smaller than the
nanostructures to be measured. Conventional techniques and data analysis
assume that the size of both heaters and temperature sensors are negligible, 
and can thus lead to unreliable results for nanostructures. We propose to 
analyze the thermal 
and electrical transport both in the
heaters/sensors and in the structures to be measured, by means of finite 
element modeling (FEM),
overcoming the approximations which are normally valid in conventional 
(macroscopic) structures.
We focus our analysis on the application of the 3$\omega$ 
method\cite{cahill-1990}, because 
the fabrication of the test structures is simpler with respect to what is
required by other techniques for the measurement of the 
thermal conductivity. 
However, our numerical method could be easily extended also to such techniques. 
The 3$\omega$ technique requires only 
the fabrication of a metal strip, which is then 
biased with an alternate current. The third harmonic of the measured voltage 
depends on the time-dependent variation of the resistance with temperature 
under the effect of the electrical current, 
which generates heat as a result of the Joule effect. The temperature 
variation depends 
on the heat dissipation in the device, which is strictly related with 
the thermal properties (thermal conductivity and specific heat) of the 
material. 
\begin{figure*}
\includegraphics[width=12 cm,keepaspectratio]{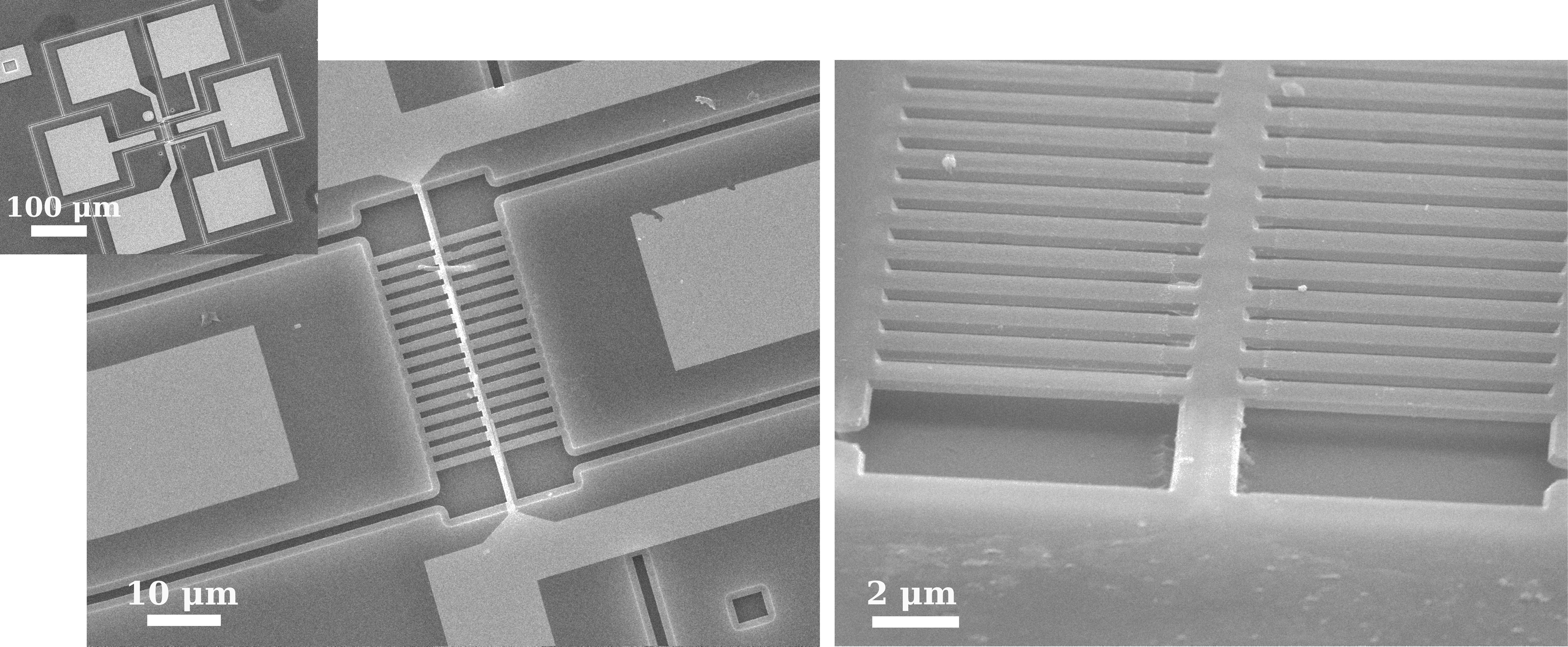}
\caption{Left panel: overall view of the suspended silicon nanoribbons, 
organized in a comb with a metal strip positioned in the middle. Right 
panel: tilted view of the device, where the suspended silicon 
nanomembranes (nanoribbons) are visible.}
\label{figura-device}
\end{figure*}
The key requirement in the 3$\omega$ technique is to define a precise model 
which 
relates the measured 
amplitude and phase of the third harmonic with the thermal properties of the material. A well assessed 
analytical model has been developed for the measurement of the thermal 
conductivity of thin films, in 
the perpendicular direction with respect to the film plane\cite{cahill-1990}. 
Analytical models for 
wires\cite{choi-2006,zhang-2001} and for suspended membranes have also been 
derived\cite{goodson-2008}. 
These models are based on the analytical solution of the heat transport 
equation, made possible at the 
price of some approximation. Three main approximations are in general 
included: 1) the electrical 
power to be dissipated is evaluated considering the value $R_0$ of the heater 
resistance at room temperature, or an average value of resistance over the 
temperature variation range; 2) the heater is considered as a one-dimensional 
heat source, so that it 
sets the boundary conditions for the solution of the heat transport equation; 
moreover, the effects due to the leads needed for supplying the electrical 
signal to the heater are neglected; 3) these models also neglect
neglect the electrical conductivity of the material under test, therefore 
they can be applied only to the thermal characterization of insulating, or 
semi-insulating, materials; moreover, they involve an assumption about the 
thermal conductivity of the heater.
All these approximations can strongly affect the results, in particular if 
very small structures are considered. We propose a different approach, 
based on the numerical solution of the thermal and electrical equations 
which describe the heat and charge transport in the structure. We then apply 
the method to the measurement of the thermal conductivity of silicon 
nanoribbons. However, the method is very
general, and, with simple modifications, it can be adapted to a large variety 
of structures.
In section II (Device fabrication and measurement setup) the fabrication of 
the device used for the
proposed characterization and the measurement set-up will be illustrated. 
In section III the numerical 
method for 3$\omega$ data reduction will be described. In section IV  a 
comparison with an analytical method will be presented.
\begin{figure*}
\includegraphics[width=7 cm,keepaspectratio]{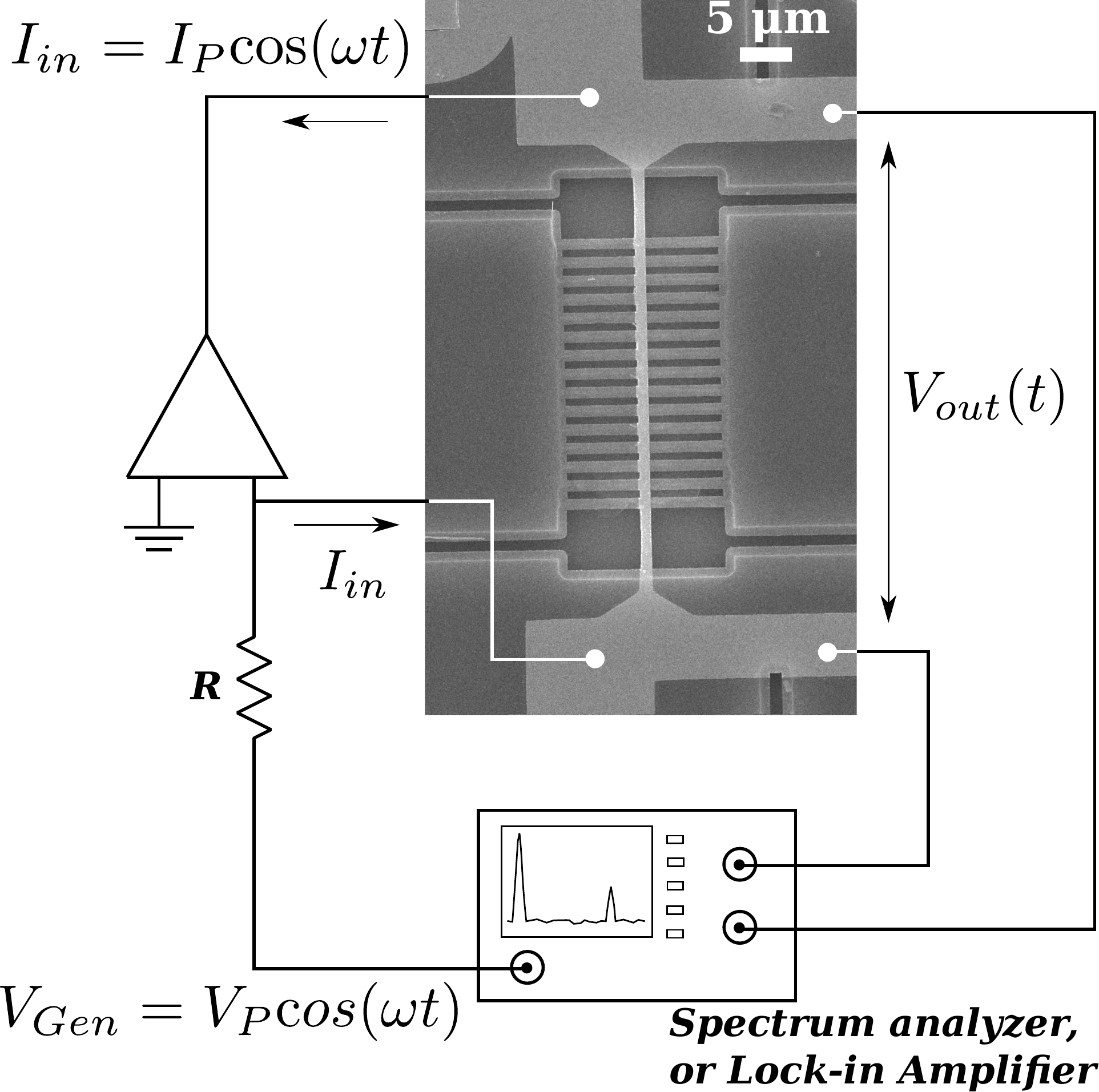}\hskip 1 truecm\includegraphics[width=8 cm,keepaspectratio]{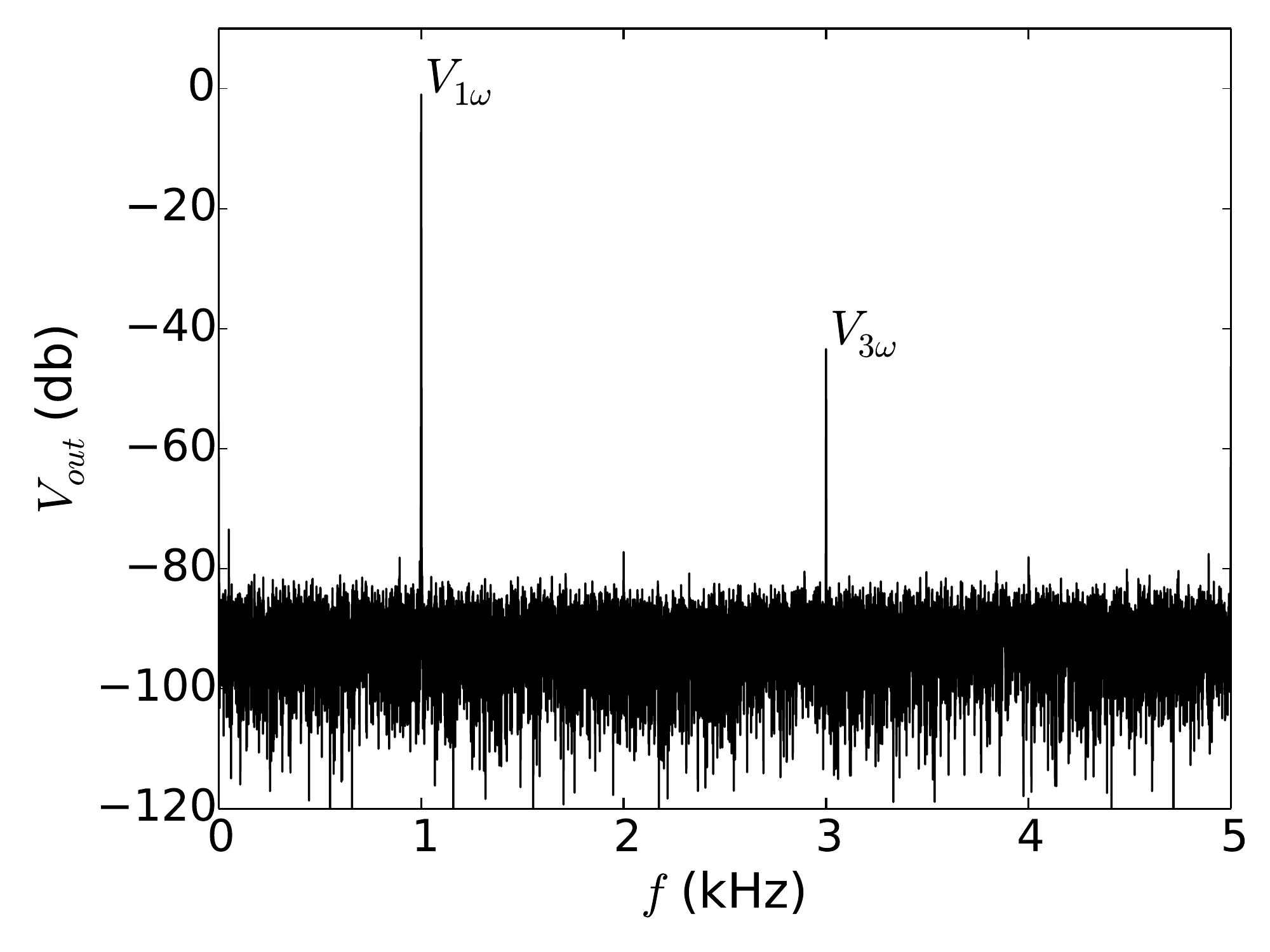}
\caption{Left panel: sketch of the measurement setup. The metal strip is 
biased with a sinusoidal current, injected through two contacts used as current 
probes. The sinusoidal current is provided by the voltage source 
of the lock-in amplifier (or of the spectrum analyzer) through a 
voltage-to-current converter. Right panel: a 
typical spectrum of the voltage, measured through the other two contacts, used 
as voltage probes.}
\label{figura-apparatus}
\end{figure*}
\section{Device fabrication and measurement setup}
Figure~\ref{figura-device} shows SEM images of the typical devices which 
were used for the 3$\omega$ measurement of thermal conductivity. 
The devices are based on monocrystalline silicon ribbons (thin nanomembranes), 
width a width $W$ between 1 and 1.2 $\mu$m, and a length $L$ between 5 and 
10 $\mu$m; the thickness
$t_h$ is 240~nm. Our aim is to measure the thermal conductivity in the film 
plane, parallel to the silicon 
surface. For this reason, the nanoribbons, arranged in a double-comb 
configuration, are suspended between the ends of the comb, as seen in the 
SEM image shown in the right panel. A metal (Gold) track is fabricated, 
exactly aligned with the center of the comb. This suspended metal resistor acts 
as the heater for the 3$\omega$ measurements. Two suspended silicon leads 
(one at the top and the other at the bottom of the comb) 
support the metal track, which is connected to the four contacts fabricated on 
the unsuspended part of the device
(see the inset in the left panel of Fig.~\ref{figura-device}).
We summarize the fabrication process, which is a modification of the one that 
we have already used for the 
fabrication of silicon nanowire devices\cite{mio-pellegrini, 
mio-nanonet-nanolet}.
We start from a Silicon On Insulator (SOI) wafer, with a top silicon layer 
260 nm thick and 
a buried oxide layer 2~$\mu$m thick. A SiO$_2$ layer 40~nm thick is 
grown at the top, and trenches
are defined by means of electron beam lithography, through PMMA resist 
exposure, development and Buffered 
Oxide Etch (BHF). The SiO$_2$ layer is then used as a mask for etching the 
top silicon layer by means
of Potassium Hydroxide (KOH etch, 35\% in water at 43$^o$ C).
The trenches are designed for the definition of the comb in the top 
silicon layer, and for providing 
electrical and thermal insulation between the different regions of 
the device.
The thickness of the top silicon layer is measured by means of Atomic Force 
Microscopy (AFM) imaging. To this end,
at first the thickness of the SiO$_2$ top layer has been measured by 
acquiring AFM images of the trenches
after the BHF etch and the resist removal by means of acetone. Then, AFM 
imaging  has been repeated after the KOH etch 
of the Si top layer, assuming that this etch stops at the buried oxide, since
it is ineffective on SiO$_2$.
In this way, the total thickness of the SiO$_2$ and of the Si top layers has 
been measured, and the correct thickness of the Si device layer has been 
obtained by difference.
Before the suspension of the nanomembranes, metal tracks and contacts have been fabricated. To this end,  
an e-beam lithographic step, precisely aligned on the silicon structures, is 
performed by using a PMMA 
resist layer. Then, a gold film 70~nm thick is deposited by means of thermal 
evaporation, and lift-off is performed in hot 
acetone. Also the exact thickness of the metal film is determined by means of 
AFM imaging.
At this point, the suspension of the silicon nanoribbons, and of the leads for the metal 
tracks, is obtained by etching the buried oxide which is under the structures 
(oxide underetching). The 
nanoribbons have a width between 1 and 1.2~$\mu$m, therefore more than 
10 minutes of BHF etch time is  
required (etch rate of about 50~nm/min) for the suspension of the comb. 
As BHF is practically ineffective on Gold, metal tracks and contacts 
are preserved. The SEM image on the left panel of 
Fig.~\ref{figura-device} shows the silicon nanoribbons organized in a 
comb configuration. Contacts for the 
electrical characterization of the central heater, designed in a four 
probe configuration, are visible
in the low magnification SEM image shown in the inset. 
Two more contacts are provided for the investigation of electrical 
transport through the silicon nanoribbons.
The SEM image shown in the right panel of Fig.~\ref{figura-device} includes 
a cross-section of the device (taken
before the fabrication of the heater).
\begin{figure*}
\includegraphics[width=12 cm,keepaspectratio]{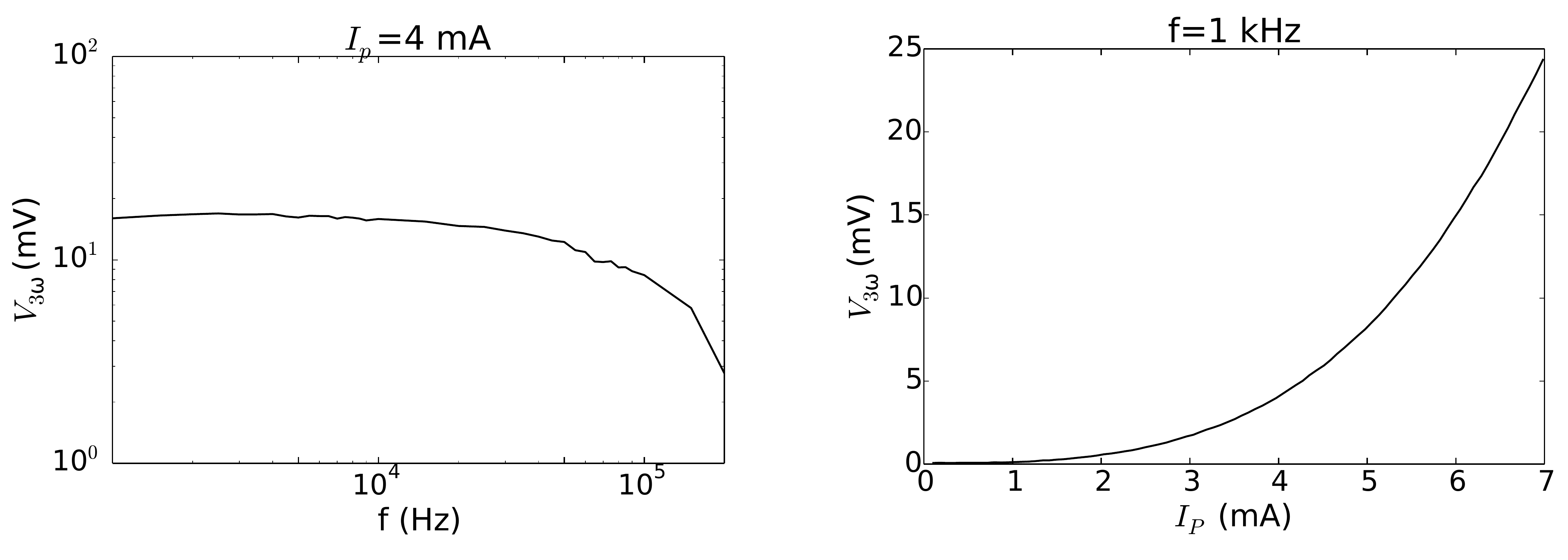}
\caption{Results of 3$\omega$ measurements on a typical device. 
Left panel: 3$\omega$ voltage amplitude as a function of the frequency, for a 
bias current $I(t)=I_P \sin(\omega t)$ with $I_P=5$ mA. 
Right panel: 3$\omega$ voltage amplitude as a function of the bias current 
$I_P$, for a constant frequency of 1~kHz.}
\label{figura-misure}
\end{figure*}
\begin{figure*}
\includegraphics[width=12 cm,keepaspectratio]{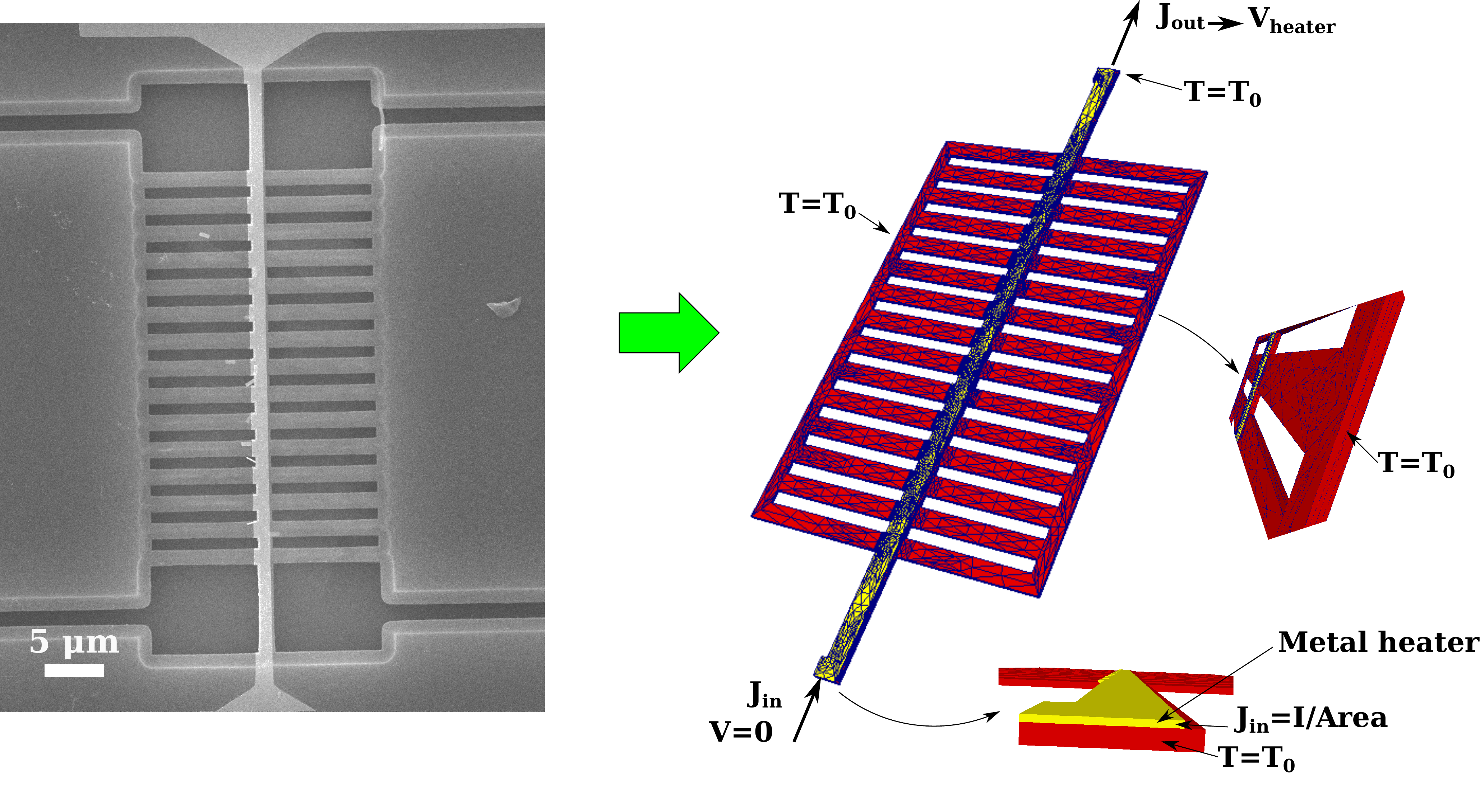}
\caption{Left panel: SEM photo of a device, similar to that already shown 
in Fig.~\ref{figura-device}. 
Right panel: a 3D model of the structure is extracted using both the 
SEM photo and the thickness measured
from AFM images. A grid is then generated and boundary conditions for 
the solution of the thermoelectric equations
are defined as shown in the figure.}
\label{figura-SEM-grid}
\end{figure*}
The left panel of Fig.~\ref{figura-apparatus} shows the experimental 
set-up used for the 3$\omega$ 
measurements. A sinusoidal current $I(t)=I_P \sin(\omega t)$ is fed between 
two (current probes) of the four contacts of the gold track fabricated in 
the middle of the comb. The voltage is 
collected through the other two contacts (voltage probes) and measured 
by means of a lock-in amplifier (Eg\&g 5302) which has a 
differential input amplifier, or by means of a digital signal analyzer. 
The sinusoidal current signal is obtained from
the internal voltage source of the lock-in amplifier (or of the digital
signal analyzer), applied to a voltage 
to current converter (transconductance amplifier), whose schematics is shown 
in Fig.~\ref{figura-apparatus}. 
The internal source provides a voltage signal $v(t)=V_P \sin(\omega t$); the 
peak amplitude of the current is $I_P=V_P/R$. A calibrated resistor 
of 470 $\Omega$ has been used for all the measurements.
The measured output impedance of the amplifier is of the order of the G$\Omega$. 
This value is far larger than the resistance of the heaters for all 
the measured devices (always in the range between 30 and 300 $\Omega$). 
The harmonic distortion of the amplifier has been tested measuring it for 
several frequencies, up to 1~MHz, by 
loading the output with a commercial 33 $\Omega$ resistor. 
The lock-in  amplifier is locked on the output voltage of its internal source,
and the amplitude and phase both of the first and of the third harmonic have 
been measured. 
The third harmonic amplitude was always below 10~$\mu$V, for first harmonic 
amplitudes as large as 5~V ($V_P=5$~V, $I_P=10.6$~mA). In the right panel of 
Fig.~\ref{figura-apparatus}, we report the spectrum 
of the output voltage when a sinusoidal current signal, with a frequency of 
1~KHz and $I_P=3$~mA, is applied to the metal heater of a typical device.  
The presence of a voltage signal (harmonic distortion) whose frequency is 
three times that of the biasing current is apparent. The measurement of the 
amplitude of this third 
harmonic distortion is the basis of the $3\omega$ technique. As the 
injected current is sinusoidal with a 
frequency $\omega$, the resulting Joule heating (proportional to the square 
of the current) has a zero frequency component 
plus a superimposed 2$\omega$ component (as a first approximation). The 
heat generated by the metal track (resistor) for the Joule effect 
depends on the biasing current $I(t)$ and on the resistance value $R(t)$: 
$P(t)=I(t)V(t)=I^2(t)R(t)$. 
This heat must be dissipated through the suspended nanoribbons/nanomembranes. 
Therefore, the temperature $T_R(t)$ of the resistor, driven by the 
instantaneous power $P(t)$, depends on the thermal conductivity $k_t$ 
and on the thermal capacity $C_V$ of the nanomembranes. A measurement of 
$T_R(t)$ allows to determine 
the thermal properties ($k_t$ and $C_V$) of the nanomembranes. The 
temperature of the heater is measured indirectly through the resistance 
$R$ of the metal track. For a reasonably small range of 
temperature variation, the relationship between $R$ and the absolute 
temperature $T$ can be considered
linear: $R(T)=R_0\left(1+\alpha(T-T_0)\right)$, where $T_0$ is a reference 
temperature, $R_0$ is the resistance 
at $T_0$ and $\alpha=(\partial R / \partial T) / R_0$ is 
a coefficient which depends on the material: 
$\alpha=0.00385$ K$^{-1}$ for Gold. Since the temperature of the 
metal heater oscillates with a frequency 2$\omega$, as the generated 
thermal power, the resistance value $R(t)=R(T_R(t))$ 
oscillates with the same frequency. Therefore, the voltage drop between 
the ends of the heater has a component with an angular 
frequency 3$\omega$, since the current varies with a frequency $\omega$. The 
amplitude and phase of the 3$\omega$ 
component are closely related with the thermal characteristics of 
the silicon nanoribbons,
because the heat generated by the metal resistor/heater must be dissipated 
through the nanomembranes. 
We performed two types of measurements as a function of frequency, one
with a constant peak amplitude $I_P$ of the bias current, 
and the other as a function of $I_P$ at a constant frequency. For 
frequencies smaller than a ``transition frequency'' $\omega_t$, the 
amplitude of the third harmonic does not depend on the frequency. 
The transition frequency is proportional to the reciprocal of the 
propagation time $\tau_t$ of the heat wave, 
which depends on the thermal diffusivity coefficient $D_t=k_t/C_V$ and on 
the length $L$ of the nanomembranes:
$\tau_t=L^2 /2 D_t$. In other words, $L$ is the penetration depth of a 
heat wave with frequency $\omega_t$.
For frequencies $\omega\ll\omega_t= 2\pi / \tau_t$, we can assume that the 
heat wave is in phase with the local temperature, therefore the 3$\omega$ 
component of the output voltage is in phase with 
the biasing current. In this case, the local temperature, and hence 
the 3$\omega$ amplitude, depend only on the thermal conductivity $k_t$. 
We performed several measurements of the 3$\omega$ amplitude as a 
function of the bias current peak amplitude $I_P$ at ``sufficiently low'' 
frequencies. Figure~\ref{figura-misure} 
reports a measurement on a silicon membrane 240~nm thick.
\section{Numerical analysis of 3$\omega$ data}
For a precise evaluation of the thermal conductivity from the 3$\omega$ 
experimental measurements, 
a sufficiently refined model of thermal transport in the considered structures 
must be applied. This is the key
point and the most difficult task involved in the application of the 
3$\omega$ technique, because the
model must take into account the thermal conductivity, the heat capacity 
of the material, the electrical conductivity, as well as the geometrical 
parameters of the device. Standard approaches for the 3$\omega$ data reduction 
make some assumptions which can lead to unreliable results, in particular 
if nanometric structures are considered.

A first approximation is made in the calculation of the generated instantaneous power $P(t)$, for which
the value $R(t)=R_0$ is considered in standard models. If the resistor is 
biased with a sinusoidal
current $I(t)=I_P\cos(\omega t)$, $P(t)$ can be written as:
\begin{eqnarray*}
P(t) &=& I^2(t) R(t) \simeq I^2(t) R_0\\
P(t) &=& {I_P^2 \over 2}\left[1+\cos(2\omega t)\right] R_0\, .
\end{eqnarray*}
The resistor temperature $T_R(t)$
has a sinusoidal variation with a frequency $2\omega$ around the average 
value $T_M$:
$T_R(t)=T_M+T_P \mathrm{cos}(2\omega t +\theta)$, where $\theta$ is the phase 
of $T_R(t)$ with
respect to $P(t)$, and $T_M-T_P< T_R(t)<T_M+T_P$. The metal track resistance 
$R(t)$ becomes:
\begin{equation}
R(T) = R_0 \left (1+\alpha\left(T_M-T_0+T_P \mathrm{cos}(2 \omega t+\theta)
\right) \right)\, .
\end{equation}
As a consequence, the measured voltage $V(t)=R(t)I(t)$ has a fundamental 
harmonic with a frequency $\omega$,
whose amplitude $V_{1 \omega}$ and phase $\beta$ depend on the temperature 
heat and on geometrical
factors, and a third harmonic component $3 \omega$. With some simple algebra, 
we obtain:
\begin{equation}
V_{out}(t)=V_{1\omega} \cos(\omega t +\beta)+ {1 \over 2} R_0 I_P \alpha 
T_P \cos(3 \omega t + \theta)\, ,
\end{equation}
where $\theta$ is the phase of the third harmonic with respect to the biasing 
current $I_P cos(\omega t)$.
The use of the value $R_0$ (at $T=T_0$) for the calculation of the generated 
instantaneous power is an approximation 
which holds if the variation of the metal track resistance $R(t)$ is very 
small. Therefore, the models for 3$\omega$ 
data reduction developed on the basis of this approximation can be used only 
when the bias current signal $I(t)$ is small. 
However, in such a case the amplitude of the third harmonic of the measured 
voltage is in turn very small with respect to that 
of the first harmonic. The overall voltage signal, including the
first and the third harmonic, is applied to the input amplifier 
of the lock-in, or of the spectrum analyzer. Thus, there is an upper limit
for the amplifier gain, which must be chosen in such way as to be compatible 
with an as linear as possible amplification of the first harmonic.
This implies that the measurement of the much smaller third harmonic will 
be affected by reduced accuracy. A trade-off must therefore be reached 
between the distortion of the first harmonic and the signal-to-noise ratio
achievable for the third harmonic component. A reasonable result was obtained
with a first harmonic drive around 0.5~V, corresponding to a third harmonic
amplitude in the millivolt range.
In this case, the assumption that 
$R_0$ can be used for the evaluation of $P(t)$ can yield unreliable results.
\begin{figure*}
\includegraphics[width=14 cm,keepaspectratio]{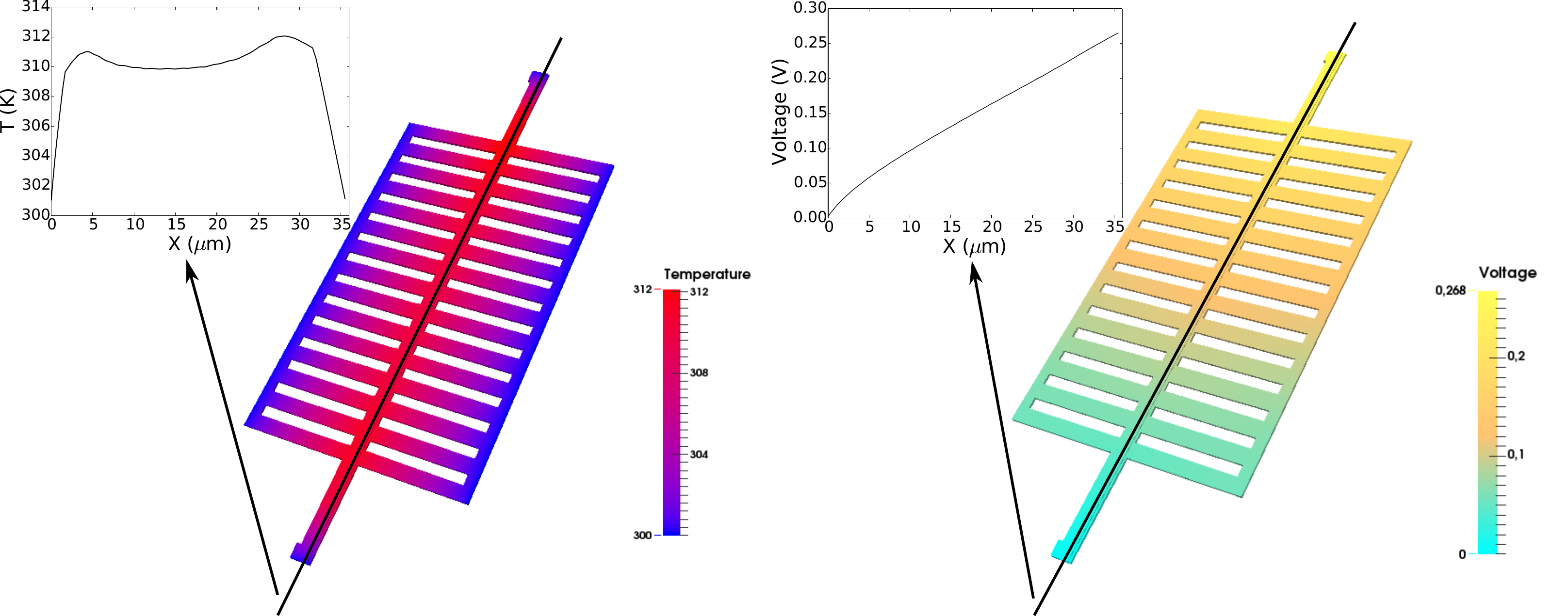}
\caption{Left panel: temperature distribution in the device measured as 
shown in Fig.~\ref{figura-SEM-grid}. 
The current was 5~mA. In the inset: the temperature in the middle axis of the 
device is reported as a function of $y$.
Right panel: Voltage distribution for the same device. The total voltage drop 
is extracted from a point very 
close to the face where the output current is imposed as a boundary condition.}
\label{figura-FEM}
\end{figure*}
The second approximation considered by standard models, consists in assuming 
that the heater is very small (of negligible extension)
with respect to the size of the structures under test. In this way, an 
analytical solution of the heat transport
equation:
\begin{equation}
{\partial T \over \partial t} = {k_t \over C_V} {\partial^2 T \over 
\partial x ^2}
\label{heat-equation}
\end{equation}
can be found because Joule heating generates a condition on one of 
the boundaries of the integration domain
(where the heater is applied). The heater sets the heat flux 
$\phi$: $\phi=-k_t\partial T /\partial x$. 
This second approximation is weak in the case of nanometric devices, 
because the width of the heater cannot be made 
very small with respect to that of the structures under test. A third 
approximation consists in taking into
account only the Joule heating of the metal track. This approximation 
is valid if materials with high 
electrical resistivity are considered. For this reason, conventional 
3$\omega$ models can be applied only 
to the measurement of the thermal conductivity of insulating, 
or semi-insulating, materials. In our case, the metal 
heater is in contact with the silicon whose thermal conductivity must be 
measured. The electrical 
conductivity of silicon is small with respect to that of metal: even 
if heavily doped silicon is considered 
(in our case $n=N_D=10^{18}$ cm$^{-3}$), its electrical conductivity 
is several order of magnitude smaller 
than that of Gold. However, the width and the thickness of the silicon 
device are larger than those of the metal 
track: for example, the thickness of the measured nanomembranes is in the 
range between 120 and 240~nm, while the 
thickness of the metal heater is always smaller than 70~nm. For this 
reason, the silicon conductivity must be 
taken into account for a correct interpretation of the experimental results.
In general, the electrical characteristics of the material can be 
determined with standard techniques, and then
the effect of the electrical conductivity can be taken into account 
considering the full thermoelectric transport 
equations. In particular, the electrical characteristics of silicon 
are known in great detail, and both its electrical
conductivity $\sigma$ and Seebeck coefficient $S$ can be determined 
by means of well-assessed semi-empirical models. However, finding 
an analytical solution which includes these semi-empirical models 
could be a very difficult task.
We therefore followed a different approach, which is based on the numerical 
solution of the thermoelectric equations. 
The technique is more demanding from the computational point of view but 
it removes all the approximations that need to be taken into account in
a practically manageable analytical solution.
At first, we obtained the exact shape of the nanoribbons and of the metal 
track (heater) from a SEM image of the device. 
The exact thickness of the metal track and of the nanomembranes was 
measured from AFM images, as explained in the 
previous section. From this information, a 3D model of the whole device 
(nanostructures and heater) was generated, as shown in 
Fig.~\ref{figura-SEM-grid}. The figure was drawn over the SEM image shown in 
the left panel of Fig.~\ref{figura-SEM-grid} by means a vector graphics 
software, and was saved in a suitable format. Then,
a python code was developed to convert the planar figure into a 
three-dimensional model, taking into account
the thicknesses of the structures. At the end, a grid generator software 
(GMSH) was used for the generation
of the mesh. On the basis of this 3D model, it was possible to solve the 
heat transport equation (Eq.~\ref{heat-equation}) 
by means of the finite element (FEM) method. However, a very significant 
computational effort would be required for fitting generic 
measurements, because transient phenomena should be taken into account 
and, hence, a very extended data set should be considered. A simpler approach 
consists in the elaboration of measurements taken in the low frequency regime, 
in which the local temperature is in phase with the heating power. In this 
case, 
the stationary thermal and electrical 
transport equations can be solved, computing voltages and temperatures 
as a function of time. The finite element 
method was used to solve the thermoelectric equations:
\begin{eqnarray*}
\vec{J} &=& \sigma \vec{\cal E} -S \sigma \nabla{T} \\
\vec{\phi} &=& ST \vec{J} - k_t \nabla{T}
\end{eqnarray*}
combined with the continuity equation for the electrical current and the 
heat equation:
\begin{eqnarray*}
\nabla{} \cdot \vec{J}(V,T) &=& 0\\
\nabla{} \cdot \vec{\phi}(V,T) &=& -\nabla{V} \cdot \vec{J}(V,T)
\end{eqnarray*}
\begin{figure*}
\includegraphics[width=14 cm,keepaspectratio]{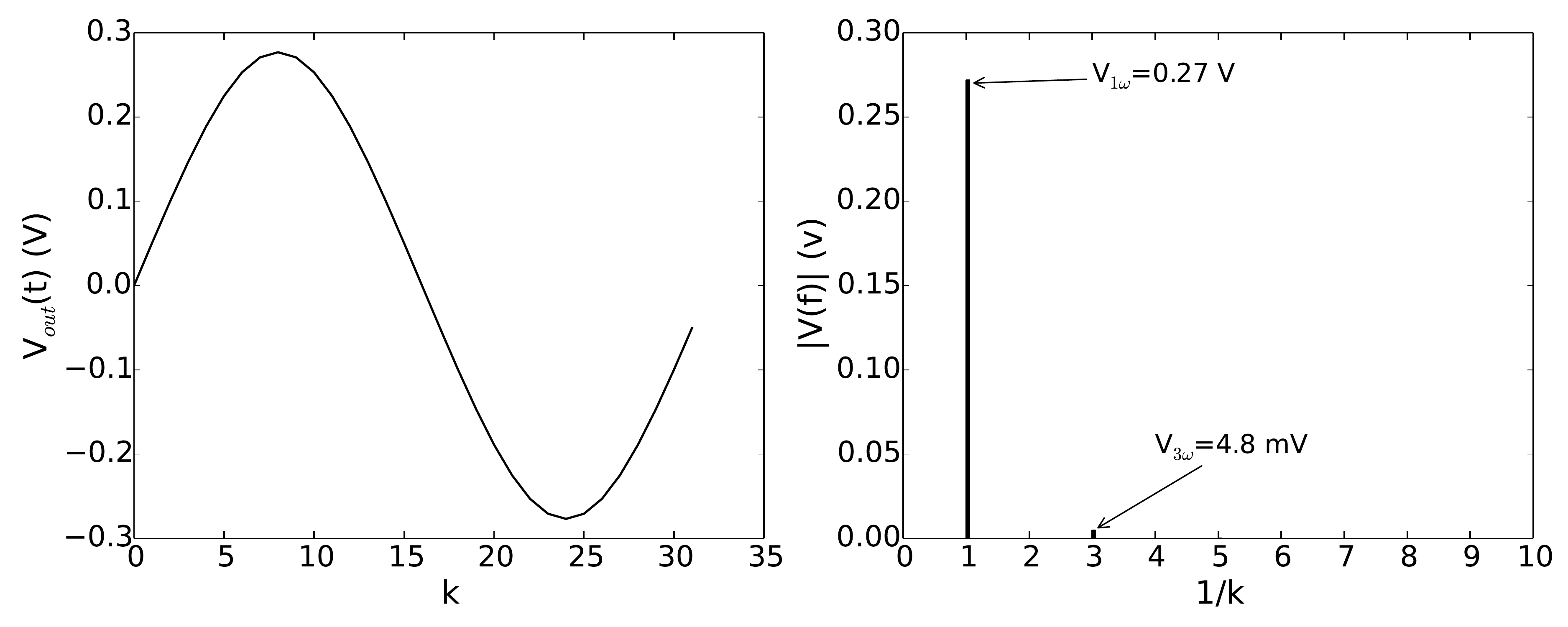}
\caption{Left panel: simulated output voltage as a function of time $t= k\Delta t$. The bias current, enforced with the boundary
conditions for the top and bottom ends of the heater, is 
$I(k \Delta t)= I_P \sin(\omega k \Delta t)$, with $I_P=5$~mA.
Right panel: Discrete Fourier Transform (evaluated
with a Fast Fourier Transform Algorithm) of the output voltage. The values 
of the first and third harmonic are indicated. }
\label{figura-FFT}
\end{figure*}
where $\sigma=\sigma(T)$ is the electrical conductivity, $S=S(T)$ is the 
Seebeck coefficient, ${\cal E}=-\nabla V$ is the 
electric field, $\phi$ is the heat flux, $k_t$ is the thermal conductivity, 
$V$ is the electrical potential, $T$ is the
absolute temperature. The heat generated by the Joule effect 
is ${\cal E} \cdot J=-\nabla{V} \cdot \vec{J}$.
The scalar fields $V$ and $T$ are the unknowns. The 3-D domain includes 
both the metal (Gold) track and the silicon
nanoribbons, which are characterized by different thermoelectric parameters 
($S$, $\sigma$ and $k_t$). 
For the metal track, Gold parameters were considered. In particular, 
$k_{t~Au}=310$ W/mK; the dependence on 
temperature of the electrical conductivity $\sigma=1/\rho$ was evaluated 
according to the linear relationship 
$\rho_{Au}(T)=\rho_{Au}(T_0)\left(1+\alpha_{Au}(T-T_0)\right)$, where 
$T_0=300$ K, $\rho_{Au}(T_0)=22.14$ n$\Omega$m, $\alpha_{Au}=0.00385$ K$^{-1}$;
the Seebeck coefficient has been assumed to be 5.1 $\mu$V/K, and it gives a 
negligible contribution to the thermoelectric transport. For the silicon 
domain (the nanoribbons), the electrical
conductivity was taken into account with the semi-empirical model of 
Arora\cite{arora}, which considers both
the effect of doping and the temperature dependence; the Slater formula 
was used to determine the 
Seebeck coefficient $S=S(T)$; the thermal conductivity $k_t$  was used as the 
fitting parameter (see below).
As boundary conditions, the room temperature $T_0$ (Dirichlet boundary 
condition) has been enforced on the sides of 
the comb and on the top and bottom faces of the silicon leads 
(see Fig.~\ref{figura-SEM-grid}). Neumann boundary 
conditions were assumed for the temperature on all the other surfaces. 
A well-defined current value was assumed in the metal 
heater through Neumann boundary conditions. For a given value of the 
current $I$, the boundary condition on each of 
the two faces at the ends of the metal strip has been $J=I/S$, where $S$ 
is the surface of the considered face. 
The potential $V$ was set to 0 on one of the two faces. The potential 
evaluated in a position close to the other 
face is the voltage drop $V_{out}$ between the ends of the metal strip; 
$V_{out}$ is the important result of the 
simulation. The numerical solution of the thermoelectric equations was 
performed with the Fenics\cite{fenics} 
Python package. Figure~\ref{figura-FEM} shows the temperature and the voltage 
evaluated for a current $I=5$~mA. In 
the insets, the profiles of the temperature and of the voltage 
in the middle of the metal heater (see figure) are shown. The amplitude of 
the third harmonic component of the 
output voltage was determined as follows. The sinusoidal 
current $I(t)=I_P \sin(\omega t)$ was sampled with a number $n$ of points 
sufficient to reasonably reproduce the waveform
(in our case $n=32$ per period): hence, 
$\Delta t=2\pi / \omega n$, $I(k)=I_P \sin(2\pi / n k)$, where $k$ is an 
integer $0\leq k<n$. For each value $I(k)$ 
of the current the output voltage $V_{out}(k)$ was determined. 
The amplitude of the third harmonic was extracted by performing a 
Discrete Fourier Transform (DFT) 
of $V_{out}(k)$. Figure~\ref{figura-FFT} shows 
the output voltage for $I_P=5$ mA (left panel) and its DFT (right panel), 
where the amplitude both of the first 
and of the third harmonic is reported. The DFT was performed by means of 
the FFT module of the numpy Python 
package. For this plot, a thermal conductivity $k_t=100$~W/mK was 
considered.
\begin{figure}
\includegraphics[width=8 cm,keepaspectratio]{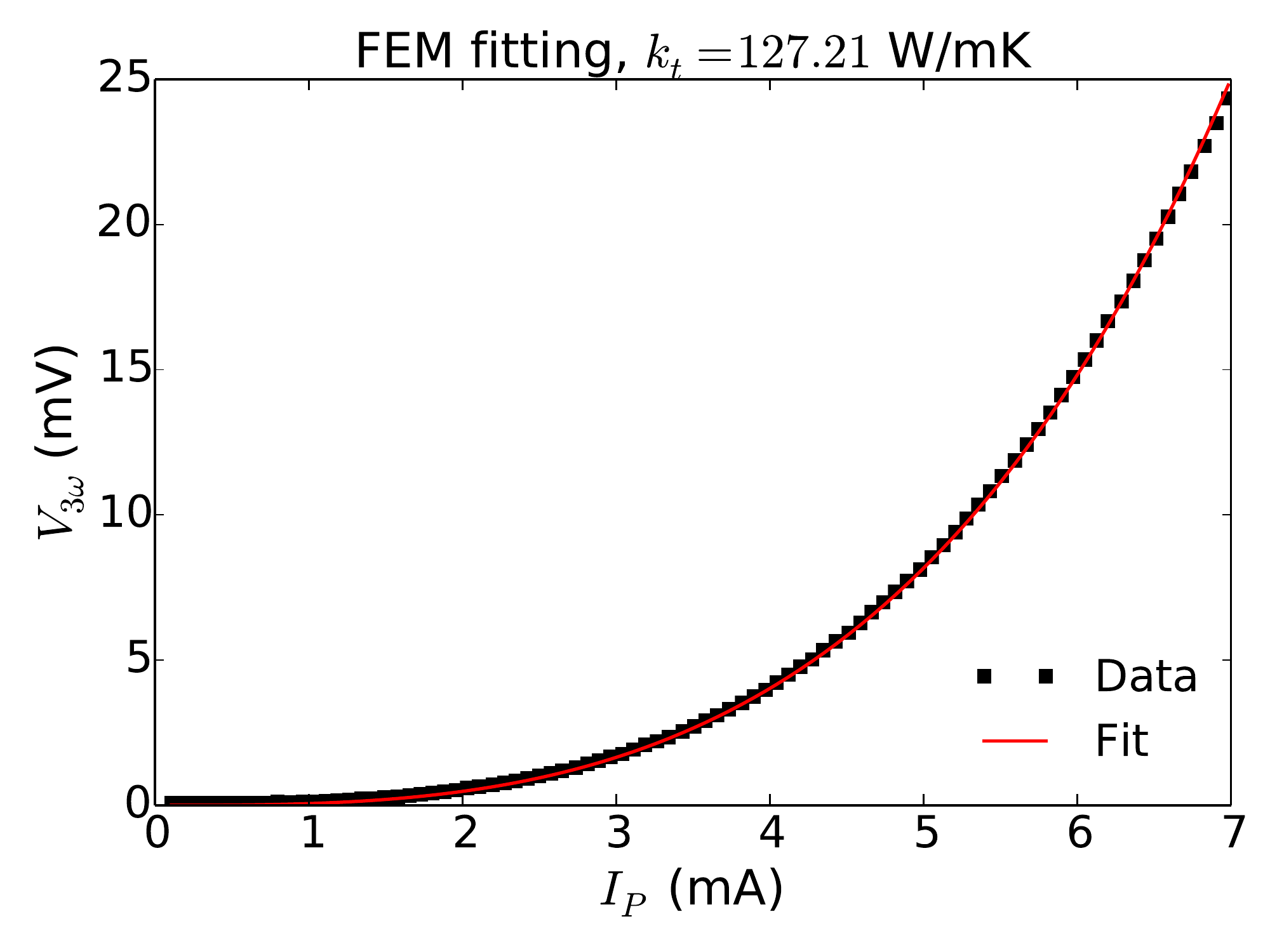}
\caption{FEM fitting of the experimental data $V_{3\omega}$ as a function 
of $I_P$ for $f=1$~kHz, shown in the right
panel of Fig.~\ref{figura-misure}. }
\label{figura-fitting-3D}
\end{figure} 
This procedure was used for fitting the experimental curve of the $V_{3\omega}$ amplitude as a function
of the peak current $I_P$, shown in Figure~\ref{figura-fitting-3D}. The thermal 
conductivity $k_t$ of the nanomembranes
was used as a fitting parameter: the amplitude of the third harmonic was
computed for all the
values of $I_P$, and $k_t$ was determined by minimizing the sum of the 
residuals obtained with respect to 
the experimental points. For the minimization of residuals, a golden section 
search algorithm\cite{numerical-recipes} 
was applied. The experimental measurements and the result of the fitting 
are shown in Fig.~\ref{figura-fitting-3D}: 
a thermal conductivity $k_t=127.21$~W/mK was obtained. This value, which 
is smaller than that of bulk
silicon ($k_t=150$~W/mK at room temperature), confirms that the thermal 
conductivity is reduced in
nanostructures, as already established by several experimental and 
theoretical studies\cite{xxx}. Our structures can be
considered as nanomembranes\cite{marconnet-2013}, with a nanometric 
thickness of $t_h-240$ nm,
and two macroscopic dimensions (1 $\mu$m wide, 5-10 $\mu$m long). 

\section{Comparison with analytical methods}
\begin{figure}
\includegraphics[width=8 cm,keepaspectratio]{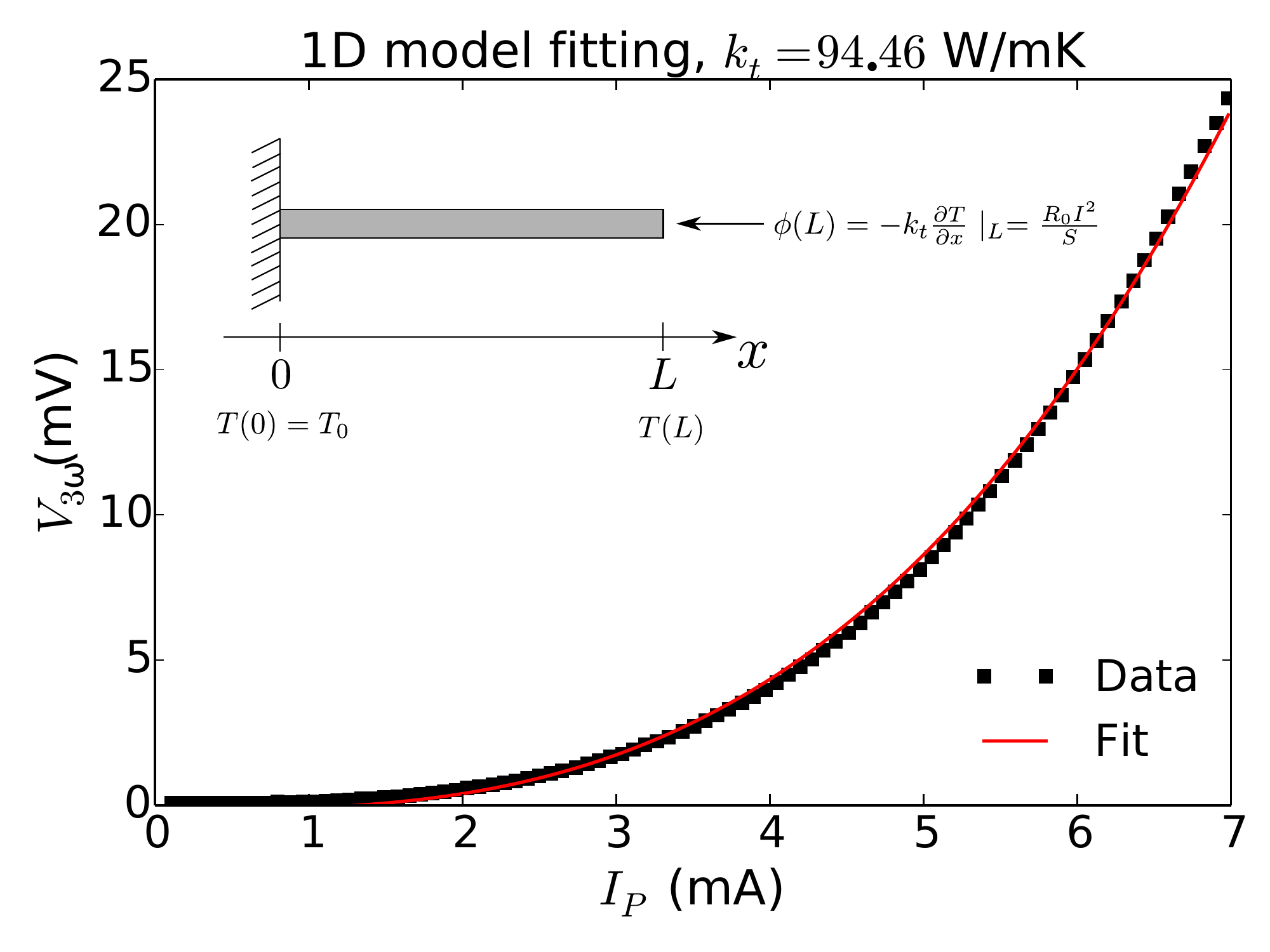}
\caption{
FEM fitting of the experimental data $V_{3\omega}$ as a function 
of $I_P$ for f=1 kHz. The experimental data, used for the fitting, are those 
shown in the right panel of Fig.~\ref{figura-misure}.
The fit is not as good as that reported 
in Fig.~\ref{figura-fitting-3D}, and the achieved thermal conductivity value 
differs more than 20\% from the one evaluated by 
means of FEM fitting.}
\label{figura-fitting-1D}
\end{figure}

In order to validate our technique, and to evaluate the influence of
the approximations needed for the development 
of an analytic solution on the final result, we used a simple 
one-dimensional model. We compare the 
results obtained by fitting the experimental measurements with this model 
to those obtained with the 3-D 
FEM model. Each silicon nanoribbon is 1 $\mu$m wide, 240 nm thick and 
7 $\mu$m long. Therefore, from 
the geometrical point of view, it can be approximated with a one-dimensional 
structure whose length is larger 
than the transverse dimensions. Our  typical structure, shown in the SEM image
of Fig.~\ref{figura-device}, 
is made up of 30 nanoribbons, plus two at the bottom and at the top of 
the comb needed for routing the electrical signals to the heater.
The nanoribbons can be considered in parallel from the thermal point 
of view, so that the whole structure can be seen as a single rod heated 
from one side with a power $R_0 I^2(t)$ 
(see the sketch in the inset of Fig.~\ref{figura-fitting-1D}).
The heat transport equation \ref{heat-equation} can be solved in 1-D, using 
a Dirichlet boundary condition 
in $x=0$ ($T(0)=T_0$), and a Neumann boundary condition in $x=L$ determined 
by Joule heating: 
\begin{equation}
\phi(x=L)=-k_t {\partial T(x) \over \partial x}\mid_{x=l}={R_0 \over S} \left(I_P \mathrm{sin}\omega t\right)^2
\end{equation}
where $S$ is the total cross section, which is obtained summing all the cross 
sections of the nanoribbons. 
The standard approximations of 1) $R(t)\simeq R_0$ for
the evaluation of Joule heating; 2) the width of the heater is negligible; and 
3) silicon has a negligible electrical conductivity, while the thermal 
conductivity of the heater is infinite, were used.
The analytical solution can be derived with some simple calculations and 
reads:
\begin{equation}
v_{3\omega}(t) = \frac{1}{2}R_{0}I_{0}\alpha\mid T_P(L)\mid cos\left(3\omega t+\angle T_{P}(L)\right)
\end{equation}
where:
\begin{eqnarray*}
T_{P}(0) &=& \frac{1}{2}\frac{I_{P}^{2}\, R_{0}}{S\, k_{t}\lambda}\,\frac{e^{-\lambda L}-e^{\lambda L}}{e^{\lambda L}+e^{-\lambda L}}\\
\lambda &=&\sqrt{\mathrm{j} \omega {C_V \over k_t}}
\end{eqnarray*}
where the imaginary unit $j$ has been used.
In the low frequency regime, for which:
\begin{equation}
\mid \lambda L \mid = \sqrt{\mathrm{j} \omega {C_V \over k_t}}~L\ll 1
\end{equation}
the expression for $v_{3\omega}(t)$ becomes:
\begin{equation}
v_{3\omega}(t)=\frac{1}{4}\alpha\,\,\frac{I_{P}^{3}\, R_{0}^2 L}{S\, k_{t}}\,\, cos\left(3\omega t+\pi\right)
\label{3omega-1d-formula}
\end{equation}
Therefore, as usual in 3$\omega$ techniques, the amplitude of the third 
harmonic turns out to be proportional to the 
third power of the current peak amplitude, $V_{3\omega}\prec I_P^3$, and to 
the square of the resistance $R_0$. The phase 
is constant and equal to $\pi$. It is easy to fit this analytical formula to 
the measurements, 
using the thermal conductivity $k_t$ as the fitting parameter. 
Figure~\ref{figura-fitting-1D} shows the fit, 
using the analytical formula\ref{3omega-1d-formula}, of the experimental data, 
whose fitting with the 3-D FEM model 
has been reported in Fig.~\ref{figura-fitting-3D}. The fit is not as good as 
that achieved with the 3-D model, and 
the value of the thermal conductivity $k_t=94.46$ W/mK is more than 20\% 
smaller.

\section{Conclusions} 
We have presented an approach to the measurement of thermal 
conductivity of silicon nanostructures based on the $3\omega$ technique with 
the support of a numerical simulation relying on an accurate thermoelectric
model. We have pointed out the sources of inaccuracy that result when 
applying to nanostructures the standard analytical approximations usually
associated with the $3\omega$ method. Our proposed approach is instead 
based on a numerical model that includes the solution, by means of a 
finite element method, of the thermoelectric equations together with the 
current continuity equation and the heat equation. An automated procedure 
has been devised to extract a geometrical model of the device from 
SEM and AFM images. The numerical model has been used to compute the time 
evolution of the voltage measured in the experiment, with a single fitting 
parameter, represented by the thermal conductivity, and in particular, to 
evaluate the amplitude of the third harmonic as a function of the amplitude 
of the injected current. By comparison with the experimental results, it 
has then been possible to obtain a good estimate of the thermal 
conductivity. The very good quality of the fitting of the experimental 
data (much better that what can be achieved with the existing approximate
analytical approaches) is evidence of the validity of the proposed numerical 
approach, which can be extended to the evaluation of the thermal conductivity 
of a wide class of nanostructures.
\bibliography{tre-omega}

\end{document}